\documentclass{article}
\usepackage{spconf,amsmath,graphicx}

\makeatletter
\renewcommand\normalsize{%
  \@setfontsize\normalsize{12pt}{14pt}%
  \abovedisplayskip 12pt plus 3pt minus 9pt%
  \abovedisplayshortskip \z@ plus 3pt%
  \belowdisplayshortskip 6.5pt plus 3.5pt minus 3pt%
  \belowdisplayskip \abovedisplayskip%
  \let\@listi\@listI}
\makeatother
\normalsize

\usepackage[margin=1in]{geometry}
\usepackage{adjustbox}

\usepackage{makecell}
\usepackage{adjustbox}
\usepackage[utf8]{inputenc}
\usepackage{graphicx}
\usepackage{amsmath}
\usepackage{amssymb}
\usepackage{hyperref}
\usepackage{setspace}
\usepackage{geometry}
\usepackage{titlesec}
\usepackage{multicol}
\usepackage{tabularx}
\usepackage{mathptmx} 
\usepackage{makecell}
\usepackage{enumitem}
\usepackage{url}
\usepackage{titlesec}
\titleclass{\subsubsubsection}{straight}[\subsubsection]

\newcounter{subsubsubsection}[subsubsection]
\renewcommand\thesubsubsubsection{\thesubsubsection.\arabic{subsubsubsection}}

\titleformat{\subsubsubsection}
  {\normalfont\normalsize\bfseries}
  {\thesubsubsubsection}{1em}{}

\titlespacing*{\subsubsubsection}
  {0pt}{3.25ex plus 1ex minus .2ex}{1em}

\setlist{nosep, leftmargin=14pt}

\usepackage{mwe} 

\title{Interpretable Motion Artificat Detection in structural Brain MRI}
\name{Author(s) Name(s)\thanks{Some author footnote.}}
\address{Author Affiliation(s)}
\name{Naveetha Nithianandam, M.Tech$^{ \dagger}$ \qquad Prabhjot Kaur, PhD
$^{\star}$ \qquad  Anil Kumar Sao, PhD$^{ \ddagger}$}

\address{%
$^{\dagger}$ Indian Institute of Technology, Madras\\
$^{\star}$ Boston Childrens Hospital, Harvard Medical School, Boston, USA\\
$^{ \ddagger}$Indian Institute of Technology, Bhilai%
}

\begin{document}
%
\maketitle
\begin{abstract}

Automated quality assessment of structural brain MRI is an important prerequisite for reliable neuroimaging analysis, but yet remains challenging due to motion artifacts and poor generalization across acquisition sites. Existing approaches based on image quality metrics (IQMs) or deep learning either requires extensive preprocessing, which incurs high computational cost, or poor generalization to unseen data. In this work, we propose a lightweight and interpretable framework for detecting motion related artifacts in T1 weighted brain MRI by extending the Discriminative Histogram of Gradient Magnitude (DHoGM) to a three dimensional space. The proposed method integrates complementary slice-level (2D) and volume-level (3D) DHoGM features through a parallel decision strategy, capturing both localized and global motion-induced degradation. Volumetric analysis is performed using overlapping 3D cuboids to achieve comprehensive spatial coverage while maintaining computational efficiency. A simple threshold-based classifier and a low parameter multilayer perceptron are used, which results in a model with only 209 trainable parameters. Our method was evaluated on the MR-ART and ABIDE datasets under both seen-site and unseen-site conditions. Experimental results demonstrate strong performance, achieving up to 94.34\% accuracy the in domain evaluation and 89\% accuracy on unseen sites, while almost completely avoiding false acceptance of poor-quality scans. Ablation studies confirms the complementary benefits of combining 2D and 3D features. Overall, the proposed approach offers an effective, efficient, and robust solution for automated MRI quality check, with strong potential for integration into large scale clinical and research workflows.

\end{abstract}
\begin{keywords}
MRI quality assessment, motion artifact detection, structural brain MRI, histogram of gradient magnitude, automated quality control, cross-site generalization
\end{keywords}

\newpage
\section{Introduction}
Automated analysis of structural brain MRI has become a cornerstone of modern neuroimaging~\cite{1, 2, 3, 4, 5}. These methods are increasingly employed for tissue and region segmentation, diagnostic support, and surgical planning~\cite{6, 7, 8, 9, 10, 11}. However, their performance remains strongly influenced by image quality, particularly when motion artifacts are present. For example, head motion is observed to lead to systematic biases artificially increasing estimates of cortical thickness \cite{6}. 

Conventional pulse sequences commonly used for acquiring structural MRI remain highly sensitive to subject motion~\cite{1}, with artifacts reported in up to 42\% of clinical scans~\cite{13}. This sensitivity results in substantial financial costs for re-scanning, and can also introduce systematic bias that artificially inflate cortical thickness estimates into downstream analyses. As a result, assessing image quality is an essential prerequisite before conducting and interpreting automated analyses.

Recent advances in motion-robust acquisition strategies such as prospective motion correction~\cite{11}, navigator-based approaches~\cite{1}, and deep learning aided reconstructions~\cite{14} have demonstrated considerable promise ~\cite{15}. Nevertheless, their widespread clinical adoption remains limited due to requirements for specialized hardware, modifications to acquisition protocols, and challenges with workflow integration \cite{15}. Importantly, these methods cannot retrospectively correct motion-related artifacts in large-scale longitudinal studies, such as the Adolescent Brain Cognitive Development Study (ABCD)  or Human Connectome Project (HCP) cohorts \cite{16, 17}. These studies already contain decades’ worth of data acquired using conventional pulse sequences~\cite{18, 19}, and thus remain constrained by the motion sensitivity inherent to those acquisitions.

Traditionally, experts classified each scan for its quality as good or bad. This process is labor-intensive, impractical at scale, and subject to inter-rater variability due to the subjective nature of visual inspection. Such inconsistencies hinder standardization and reproducibility, both of which are essential in clinical and research settings. There is, therefore, a growing need for automated, standardized quality analysis mechanisms that reduce human effort while matching expert level accuracy. These systems ensure that only high quality, artifact free images are included in diagnostic workflows and research.

The task of automated image quality assessment has been extensively explored in the literature through various methodologies. A widely adopted strategy involves the extraction of Image Quality Metrics (IQMs)~\cite{45}
, which quantitatively characterize different aspects of image degradation, such as motion artifacts, noise and blurring. These metrics are used as feature representations for traditional machine learning classifiers. Despite the interpretability of IQMs, the computation of these outcomes is challenged with heavy pre-processing of given images leading to long run time of these pipelines.  

With advent of DL based methods, it has enabled automated quality assessment with minimal preprocessing providing fast alternative to the IQM based machine learning methods. However, the performances of such methods have been found to be equivalent to that of previous methods~\cite{7, 21}. Additionally, the
generalization of both conventional methods and DL methods is
reported to be poor for unseen site \cite{21}. This underscores the need for methods which are computational inexpensive yet generalizes well to images from new sites, is the focus of this work.


In our previous work~\cite{20}, we introduced a novel IQM for brain MRI, Discriminative Histogram of Gradient Magnitude (DHoGM) which requires minimal pre-processing. We demonstrated that DHoGM effectively captures the distributional differences in image gradients induced by motion, thereby providing strong discriminative power for motion-related artifacts. These features were used to train a two-layer fully connected network for motion detection. While the method achieved state-of-the-art generalization, it had notable limitations: it relied on selecting and processing a subset of 2D slices, and thus failed to leverage the full volumetric information and cross-orientation interactions present in 3D MR data.

To overcome the mentioned shortcomings, the present work extends our earlier 2D-slice-based framework to a 3D formulation. Specifically, we propose a parallel learning strategy that integrates both slice-level DHoGM features (extracted from 2D slices) and volume-level DHoGM features (computed from the entire 3D volume) to predict the quality score of the  volume. 

We extend our earlier work with the following contributions:
(i) novel strategy to assess quality of volume based on 3D-cuboids thereby increasing spatial coverage and accuracy, 
    (ii) expanded the study to analyze the effect of noise onto DHoGM,
    (iii) extensive evaluation by including multiple datasets, 
    (iv) statistical evaluation on method outcomes derived from the parallel feature processing pathway, which provides interpretability feature towards model’s decisions.

The paper is organized as follows: Section~\ref{sec:methods} describes the datasets used in this work, and the proposed method. Results section presents qualitative and quantitative evaluations of the proposed model. (iii)The discussion reviews our methods' technical strengths and simplicity, compares them with existing approaches, highlights key innovations, and outlines future research directions.

\section{Material and Methods}
\label{sec:methods}
To predict motion in MRI volumes, four key components are required:
(i) a dataset of MRI volumes labeled with quality scores;
(ii) a multi-site attribute in dataset to evaluate the method’s generalizability;
(iii) a standardized workflow or pipeline for preprocessing the input volumes; and
(iv) a modeling framework that extracts features and outputs corresponding quality scores. In the following subsections, we provide detailed descriptions of each of these requirements

\subsection{Dataset}

In this study, we utilize two publicly accessible T1-weighted (T1w) brain MRI datasets: MR-ART, specifically acquired to characterize motion effects, and ABIDE, a heterogeneous multi-site dataset well-suited for evaluating generalizability across diverse acquisitions. Since both datasets are publicly available, our experiments can be readily reproduced, thereby supporting transparency and reproducibility.

\begin{itemize}
    \item \textbf{MR-ART Dataset :} The MR-ART dataset comprises 3D T1w brain MRI volumes from 148 healthy adult participants, with a median age of 25.16 years. Each scan was acquired under controlled motion conditions categorized as no motion, slight motion, and severe motion. Expert raters assigned a motion quality score \( S_i \in \{1, 2, 3\} \) to each volume, corresponding to the level of artifact severity. For this study, we binarized these labels into two classes: ``Good Quality'' (\( C_i = 1 \), corresponding to \( S_i = 1 \)) and ``Poor Quality'' (\( C_i = 2 \), corresponding to \( S_i = 2 \) or \( 3 \)).
    \item \textbf{ABIDE Dataset :} The ABIDE dataset consists of 1,112 T1w structural brain MRIs from individuals aged 7–64 years. Each scan was independently rated for quality by three experts on a three-point scale: -1 (poor), 0 (borderline), and 1 (good). For this study, a scan was labeled ``Good Quality'' only if all raters agreed with a score of 1. All other scans were categorized as ``Poor Quality.''
\end{itemize}

\subsection{Pre-processing of Data}

All MRI volumes were skull-stripped using the BET tool in FSL (v6.0.7.13). The resulting brain masks were used to extract brain-only voxels. Intensity normalization was performed using percentile normalization to account for inter-scan variability. Volumes were resized or padded to a uniform shape of \textbf{V} = \(192 \times 256 \times 256\) to ensure consistent input dimensions across both datasets. These steps were applied uniformly to both training and testing cohorts.

\subsection{Theory}
We provide a brief overview of the computation of DHoGM features below (please refer to~\cite{20} for more details). This is followed by an extension to our proposed 3D analysis–based metric, along with a demonstration of its discriminative ability for detecting motion and noise artifacts.
\begin{itemize}
    \item {\textbf{DHoGM (Discriminative Histogram of Gradient Magnitude): }}

The natural distribution of gradient magnitudes can be calculated as, 
\begin{equation}
|\nabla f(x,y)| = \sqrt{
\left( \frac{\partial f(x,y)}{\partial x} \right)^2 + 
\left( \frac{\partial f(x,y)}{\partial y} \right)^2 
}.
\end{equation}

The distribution of gradient magnitude is proved to be effected with motion~\cite{3}. We quantified this distortion using a slope-based measure across the initial bins of the gradient histogram. Our proposed DHoGM counts pixels in the initial non-zero HoGM bins and estimate the slope using the following formula, 

\begin{equation}
D = \frac{\sum_{n=2}^{5} \big( h[n] - h[n-1] \big)}{h[1]}.
\end{equation}

 For each MRI, DHoGM is computed from selected middle slices to form a 3D feature vector, which is classified, and final labels are obtained by aggregating slice-level predictions. The classification depicted in  Fig.~\ref{fig:2D_flowchart} has been published in \cite{20}. The parameter D increases in the presence of motion and remains low for motion free slices.

\begin{figure}[h!]
\centering
\includegraphics[width=0.8\textwidth]{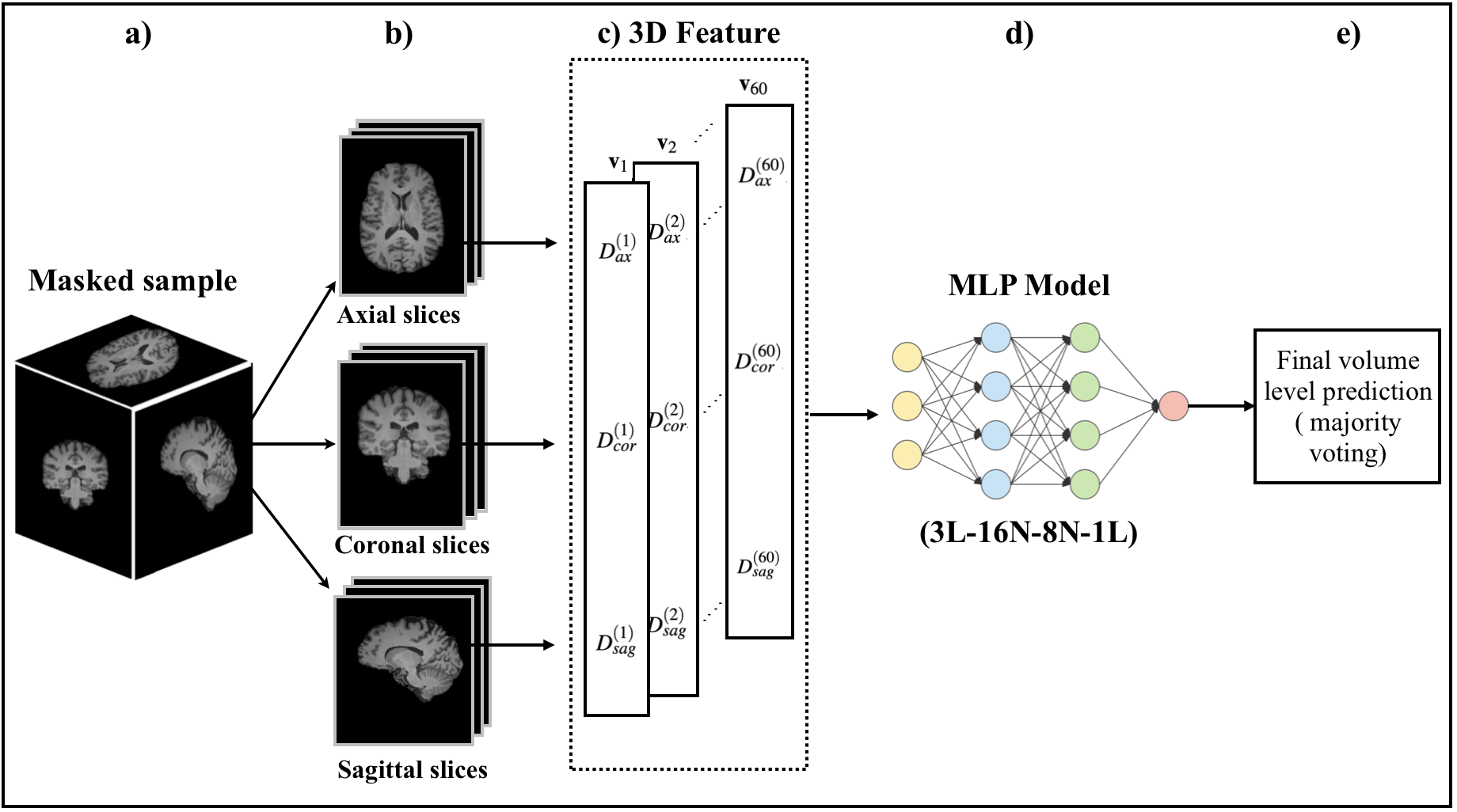} 
\caption{Sixty 2D slices per orientation are extracted from the masked sample and the 3D feature vector that is built is used to train the MLP model based on which the final predicted label is given.}
\label{fig:2D_flowchart}
\end{figure}

\item{\textbf{Proposed Metric - 3D HoGM}}
Complete volumetric coverage requires traversing the super-set of all combinations of slices across axial, coronal, and sagittal planes. Mathematically, this is represented as:
\begin{equation}
\textbf{v}_\text{ijk} = \left\{ D_\text{ax}^{i}, D_\text{cor}^{j}, D_\text{sag}^{k} \right\} \quad \text{for } i, j, k \in \{1, 2, \dots, 60\}.
\end{equation} 
In contrast, our previous 2D-based method restricted spatial coverage by imposing i=j=k in Eq.~(3), thereby simplifying the formulation. However, while the 3D extension provides full volumetric coverage, it remains computationally expensive due to the large feature vector size (e.g., approximately 216,000 feature vectors per volume at a spatial resolution of 256×256×256), which diverges from our objective of developing a fast and efficient model. To address this computational challenge, we reduced the effective volume size by extracting 3D patches (cuboids). For each 3D patch, a histogram of gradient magnitudes (3D-HoGM) was then computed, as illustrated in Fig.~\ref{fig:C1_HoGM_3D} and Fig.~\ref{fig:C2_HoGM_3D}.

\begin{figure}[h!]
\centering
\includegraphics[width=0.8\textwidth]{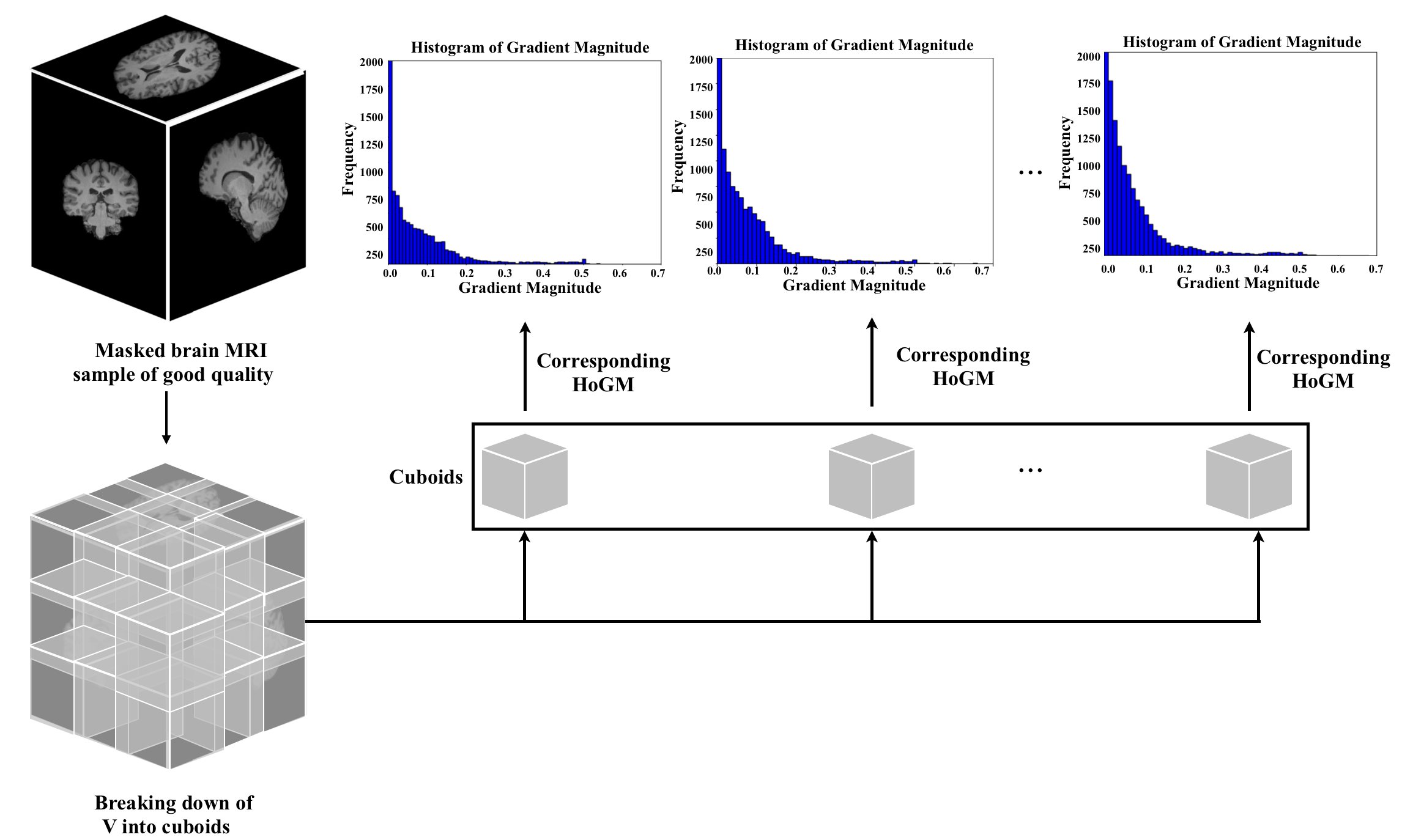} 
\caption{Monotonic nature of 3D-HoGM of cuboids of good quality samples.}
\label{fig:C1_HoGM_3D}
\end{figure}

\begin{figure}[h!]
\centering
\includegraphics[width=0.8\textwidth]{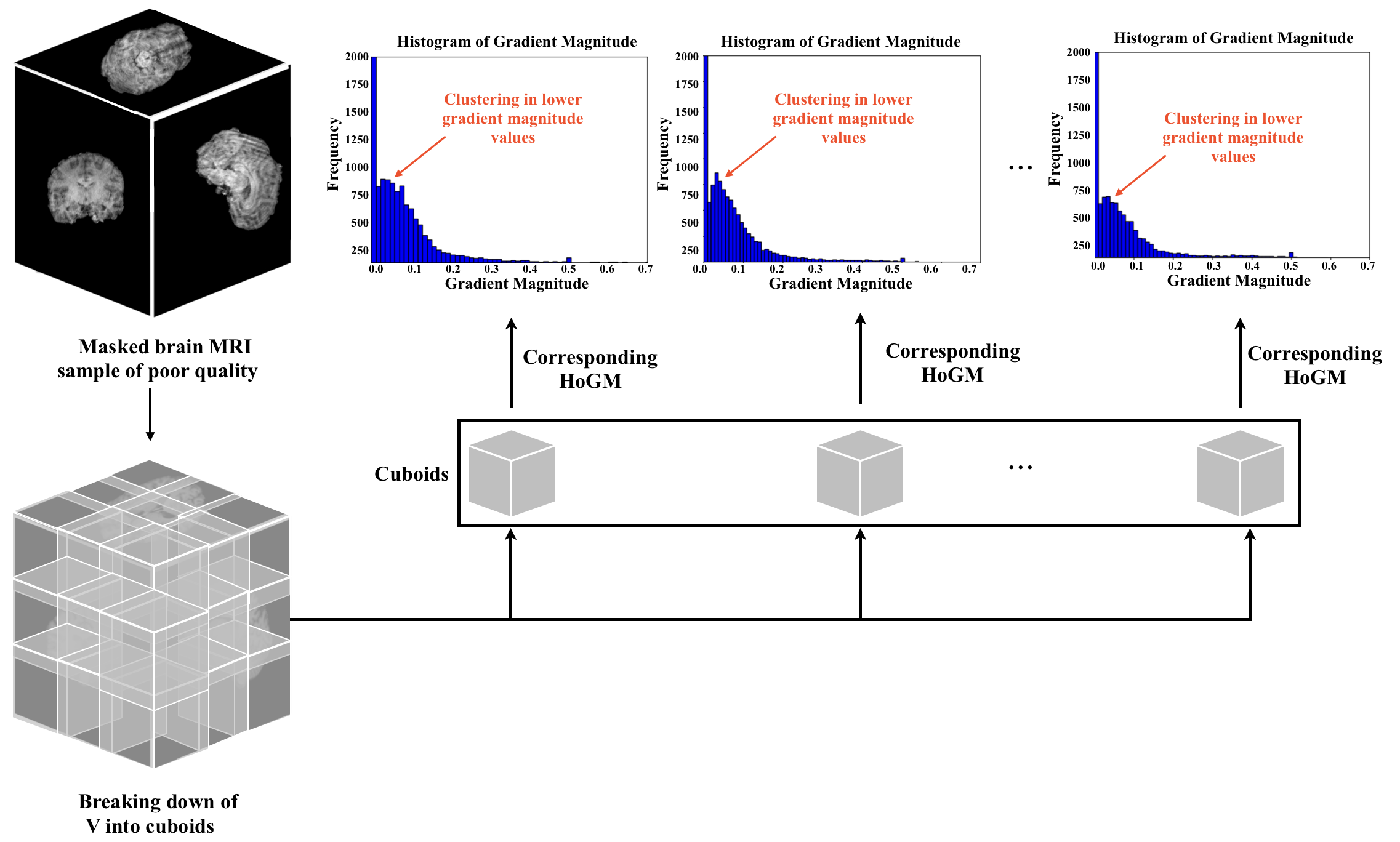} 
\caption{Clustering of lower gradient magnitude values of 3D-HoGM of cuboids of poor quality samples.}
\label{fig:C2_HoGM_3D}
\end{figure}

\item{\textbf{Discriminative Ability of 3DHoGM for Motion Detection: }}
To demonstrate the discriminative ability of the proposed metric, we conducted an experiment using a randomly selected subject who was scanned twice, once with motion and once without. The 3D-HoGM features were computed for 3D patches from each acquisition, and the corresponding patches were then compared to assess distortions in the 3D-HoGM features induced by motion. In clean images, 3D-HoGMs exhibit a monotonically decreasing trend, while motion-corrupted volumes show flatter curves and increased low-to-mid-range gradients due to blurring and artificial enhancement of smooth regions,  which exhibits characteristics analogous to those of the 2D-HoGM. The nature of 3D-HoGM for the cuboids of good quality sample is represented in Fig.\ref{fig:C1_HoGM_3D} and the nature of 3D-HoGM for the cuboids of poor quality sample is represented in Fig.\ref{fig:C2_HoGM_3D}. This allows the discriminative feature to be effectively applied in the 3D context as well.

 \end{itemize}

\subsection{Proposed Workflow: Image Quality Assessment Using 3D-DHOGM: }
A method was developed to combine quality scores of different 3D patches of a volume to predict the quality score for the input volume. 
This section elaborates on detailing sequence of processing steps including computation of DHoGM features, and proposed scoring strategy designed to combine outputs effectively. We provide a step-by-step discussion of each component in the pipeline.
\begin{itemize}
    \item {\textbf{Step 1- Extract 3D Cubes from MRI Volume: }}

The full brain volume \( \mathbf{V} \) is subdivided into cuboids of size \(96 \times 128 \times 128\) with a 20\% overlap between adjacent blocks as in Fig.\ref{fig:3D_flowchart}. This specific cuboid dimension is selected as it is sufficiently large to capture meaningful anatomical features while being constrained enough to focus on specific regions affected by motion-induced distortions. The overlap is incorporated to ensure that important structural information near the cuboid boundaries is preserved, thereby enhancing continuity and consistency in feature extraction. This partitioning strategy results in 27 cuboids per subject, offering a well-balanced trade-off between anatomical resolution and computational efficiency.  Each cuboid is processed independently to extract 3D-HoGM-based features. The mean of \( D_{3D} \) values across all cuboids forms a single scalar feature \( D_{\text{final}} \) used for classification.

\item{\textbf{Step 2- Compute Proposed-Metric Features for Each Cube: }}

Formally, in the proposed approach explained in figure\ref{fig:3D_flowchart}, the brain MRI volume is partitioned into a set of cuboidal regions of interest (ROIs), each comprising a subset of voxels. For each cuboid, the novel discriminative feature, denoted as \(D_{3D}\), is computed. These features are then aggregated to yield a single one-dimensional feature value per sample. To perform classification, the distribution of these feature vector values are analyzed across the training samples, and a decision boundary is established using a threshold-based method. This threshold is subsequently applied to test samples to predict their final class labels. 
\begin{figure}[h!]
\centering
\includegraphics[width=0.8\textwidth]{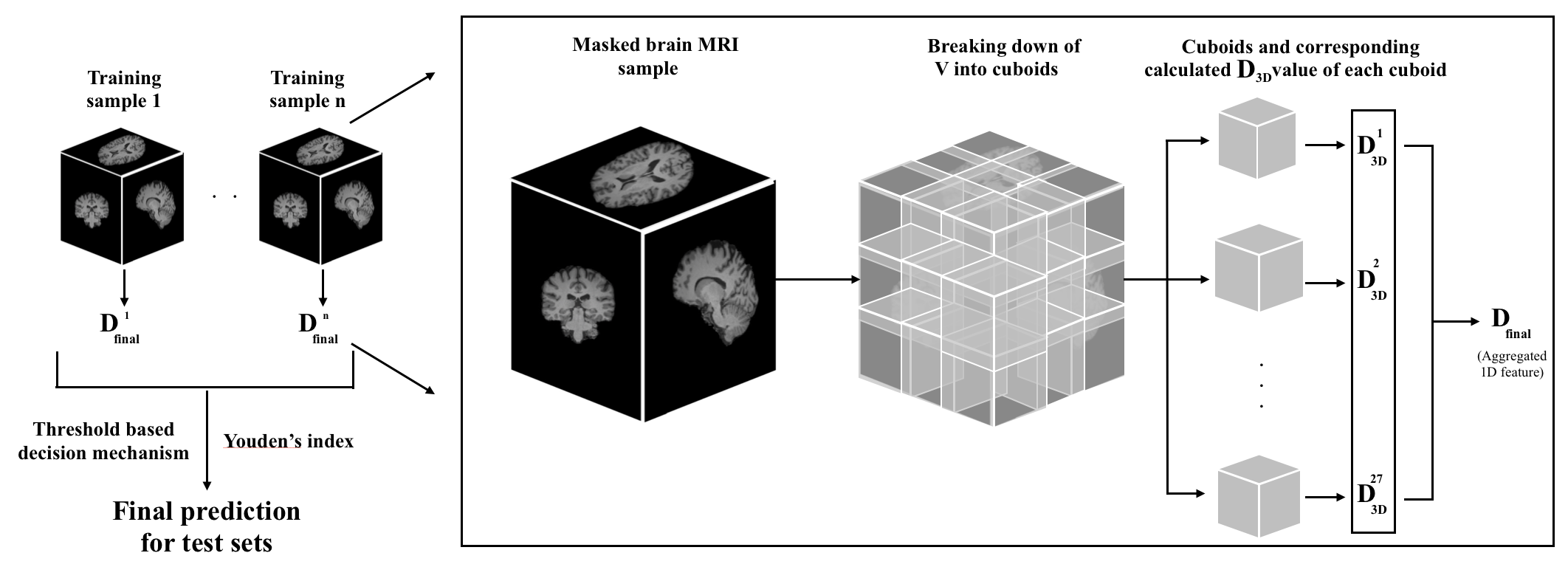} 
\caption{ For each sample, the \(D_{final}\) value is calculated by taking aggregate of \(D_{3D}\) values across all cuboids within that sample. A threshold-based decision mechanism is then used to classify the samples, where the threshold is determined based on the distribution of \(D_{final}\) values.}
\label{fig:3D_flowchart}
\end{figure}

For all the voxels in each cuboid, we compute 3D gradient magnitudes using the formula:

\begin{equation}
|\nabla f(x,y,z)| = \sqrt{\sum_{i=1}^{3} \left( \frac{ f(x,y,z)}{ d_i} \right)^2}
\end{equation} where \(   {\frac{ f(x,y,z)}{ d_x}}  \), \( {\frac{ f(x,y,z)}{ d_y}}  \), \(   {\frac{ f(x,y,z)}{ d_z}}  \) representing the gradient along  the \( x \)-direction, \( y \)-direction, \( z \)-direction respectively and then plot the 3D-HoGM per cuboid. In the volumetric approach, we define the corresponding discriminative feature DHoGM as \( D_{3D} \) which is the normalized slope across early histogram bins:
\begin{equation}
D_{3D} = \frac{\sum_{n=1}^{5} (h[n] - h[n-1])}{h[1]}
\end{equation} where \( h[n] \) denotes the bin count in the 3D-HoGM. This results in a robust summary of local structural degradation due to motion.


\item{\textbf{Step 3 - Assign Quality Scores to each sample: }}

To analyze the distribution of the one-dimensional feature \(D_{final}\) across all samples, a histogram was plotted for visualization. The separation between “Good Quality” training samples and “Poor Quality” training samples appeared to be well-defined. This observation motivated the adoption of a simple, yet effective, one-dimensional classification strategy based on thresholding. To implement this threshold-based classification, an optimal decision boundary needed to be determined. The optimal threshold in this work is derived using Youden’s Index\cite{42}.
Once the optimal threshold is identified, each sample’s final feature value \(D_{\text{final}}^i\) is compared against this threshold to assign a class label. If the \(D_{\text{final}}^i\) value of a sample is lesser than threshold, then the sample is classified as Class 1, else Class 2. Mathematically, 
\begin{equation}
C_{\text{3D}}^i = 1 \text{ if } D_{\text{final}}^i < t^*, \text{ else } 2 \quad \text{where } C_{\text{3D}}^i \text{ denotes the predicted class for sample } i
\end{equation}


\item{\textbf{Step 4- Final Prediction based on parallel approach and confidence score calculation: }}

In parallel with the volumetric approach, we train a simple multilayer perceptron (MLP) using slice-wise DHoGM features. For each of the 60 selected slices across three orientations, a 3D vector is constructed and used as input to the MLP. The model outputs per-slice predictions, which are aggregated via majority voting to assign a final label as explained in Fig.\ref{fig:2D_flowchart}. 
This dual-path classification strategy enhances both interpretability and robustness. The final classification decision is derived by integrating the predictions from both models. The fusion strategy behaves like an AND operator, where the final label is assigned as Good Quality only if both models agree on that prediction. Here, class label \(1\) indicates Good Quality, and label \(2\) represents Poor Quality. This decision rule is formally defined as:
\begin{equation}
C_{\text{final}} = 1 \text{ if } (C_{\text{2D}} = 1 \text{ and } C_{\text{3D}} = 1), \text{ else } 2 \quad \text{where } C_{\text{final}} \text{ is the final predicted class}
\end{equation}

If \( C_{\text{final}} = 1 \), the corresponding confidence score is denoted as \( P_1 \). This score is computed by averaging the confidence values \( P_{2D,C1} \) and \( P_{3D,C1} \), which represent the confidence of the sample being classified as Class 1 by the localized model \( C_{2D} \) and the globalised model \( C_{3D} \), respectively. Conversely, if the final predicted label is \( C_{\text{final}} = 2 \) (Poor Quality), the confidence score is denoted as \( P_2 \), which is the average of the confidence values \( P_{2D,C2} \) and \( P_{3D,C2} \), representing the confidence of classification as Class 2 by the respective models. This approach ensures that the final confidence score reflects the consensus of both models. The confidence scores \( P_1 \) and \( P_2 \) are defined as:
\begin{equation}
P_1 = \frac{P_{2D,C1} + P_{3D,C1}}{2}, \quad
P_2 = \frac{P_{2D,C2} + P_{3D,C2}}{2} 
\quad \text{for } P_{d, C_k},\ d \in \{2D, 3D\},\ k \in \{1,2\}
\end{equation} where \( P_{d, C_k} \) represents the confidence score assigned by the HoGM model in dimension \( d \) for class \( k \). This formulation provides a coherent and interpretable measure of classification certainty by integrating both local and global perspectives.

\end{itemize}

\subsection{Experimental Settings}

The methodology described in the preceding section was applied within the experimental framework encompassing data preprocessing and the construction of training and testing cohorts. The resulting approach was subsequently assessed following the evaluation protocol detailed in the subsequent section.

\subsubsection{Training and Test Cohort Construction}

The proposed method was evaluated on 200 samples using a stratified 50\%-50\% train-test split under two settings: (i) seen-site evaluation, conducted separately on MR-ART and ABIDE, and (ii) unseen-site evaluation, where the model was trained on MR-ART and tested on the unseen ABIDE dataset. All splits were maintained class balance across folds.

\subsubsection{Evaluation Strategy}
To assess the performance and robustness of our proposed method, we conducted extensive evaluation under the following experimental configurations: 
\begin{enumerate}
    \item Seen site evaluation - In this strategy, the model is trained and tested using samples originating from the same site. Consequently, the model is exposed to the characteristic patterns and variations specific to that site during training, providing it prior knowledge of the site-specific data distribution when evaluated on the test samples from the same source.
    \item Unseen site  evaluation - Complementary to the seen site evaluation, it is critical to assess model performance on unseen site data. In this approach, the model is trained on samples from one site and tested on samples from an entirely distinct, previously unseen site. This evaluation ensures that the model demonstrates robustness and generalizability across different data sources and remains capable of effectively handling any novel sample types it has not encountered during training.

    \item Comparison with Existing Methods -  The evaluation framework also includes a comparative analysis against established methods reported in the literature. This comparison serves to benchmark the performance of these approaches and to position our model’s effectiveness relative to the current state-of-the-art.
    \item Ablation Study - To further understand the contributions of individual components within our framework, we conducted an ablation study focused on the impact of different feature extraction strategies. This analysis evaluates the performance of each strategy independently as well as their combined effect on the overall system accuracy.
    \item Runtime and Efficiency - This evaluation provides critical insights into the computational resource requirements and processing time per sample for our model. Such metrics enable a meaningful comparison with existing methods in the literature, thus positioning the efficiency and feasibility of our approach within the existing literature.

\end{enumerate}

\subsubsection{Performance Metrics}

Model evaluation was carried out using standard classification metrics: accuracy, precision, recall, and F1-score. Accuracy reflects the overall proportion of correct predictions, precision measures the correctness of positive predictions, recall assesses the completeness of positive case detection, and the F1-score 
balances precision and recall through their harmonic mean. The formal definitions of these metrics follow the standard formulations as described in \cite{22}.

\section{Results}

This section presents the quantitative and qualitative evaluation of the proposed method. We compare its performance across multiple experimental settings and benchmark it against existing methods, analyzing accuracy, precision, recall, F1-score, and generalizability. All metrics are averaged across five folds for cross-validation experiments. Since the test sets are considered to be balanced for both classes, the accuracy mentioned here is the same as balanced accuracy.  

\begin{table}[h]
\centering
\caption{Performance of the proposed method across different datasets.}
\label{tab:quant_results}
\begin{tabular}{|l|c|c|c|c|c|}
\hline
\textbf{Dataset} & \textbf{Setting} & \textbf{Accuracy} & \textbf{Precision} & \textbf{Recall} & \textbf{F1-Score} \\
\hline
MR-ART & In-domain (5-fold) & 94.34\% & 94.16\% & 96.34\% & 95.25\% \\
ABIDE & Cross-domain & 89.00\% & 91.00\% & 89.00\% & 89.00\% \\
ABIDE & In-domain (5-fold) & 90.00\% & 90.50\% & 88.80\% & 89.67\% \\
\hline
\end{tabular}
\end{table}

\begin{enumerate}

\item{Seen-Site evaluation : }
The details are displayed in table~\ref{tab:quant_results}. Our proposed work when trained on
\begin{enumerate}
    \item MR-ART dataset achieves an average accuracy of 94.34\% with an highest accuracy of 95\%. The corresponding precision, recall and F1-score for 50 samples each per class for the test set is 94.16\%, 96.34\%, 95.25\%.
    \item ABIDE dataset with 50 samples per class for testing, fetched an average accuracy of 90.00\% for a 5-fold cross validation set with 92\% highest accuracy. The precision, recall and F1-score are 90.5\%, 88.8\% and 89.67\% respectively.
\end{enumerate}


\item{Unseen-site evaluation : }
In the second experimental setup, our model achieved an overall accuracy of 89\%, with an average precision of 91\%, recall of 89\%, and F1-score of 89\%, as detailed in Table \ref{tab:quant_results}.  A particularly noteworthy observation is that no samples of "Poor quality" were misclassified as "Good quality", thereby eliminating the risk of false positives, an essential consideration in clinical and medical imaging processes. All observed misclassifications involved "Good quality" samples being incorrectly predicted as "Poor quality." This suggests that the model is highly sensitive to subtle artifacts or degradations, even those that are often considered acceptable by experienced technicians. Such sensitivity ensures a stringent quality control standard. Furthermore, the model demonstrated high confidence in its correct predictions, with "Poor quality" samples exhibiting an average confidence score of 0.95. In contrast, the misclassified "Good quality" samples had an average confidence score of approximately 0.50, suggesting a clear separation in model certainty across quality classes.


\item{Comparison with Existing Methods :} According to Table \ref{tab:comparison}, \cite{1} trained their model using four datasets, UK-bank, OASIS-3, MR-ART and in-house data for the classification of '0' (clinically usable images with a quality score of either 1 or 2) or '1' (clinically unusable images with a quality score of 3). They achieved a highest balanced accuracy of 94.41\%, with an accuracy of 92.94\%, sensitivity of 96.43\% and specificity of 92.39\%. Exclusively for MR-ART dataset, they have achieved an accuracy of 84.17\% and recall of 97.91\%. With a good accuracy achieved by their method, it is to be noted that 220,457 parameters were used for the CNN model that fetched their highest accuracy which is computationally heavy for approximately 1600 and 400 brain MRI samples for training and testing respectively. Also, the ratio of samples per class of '0' to '1' was 25:4 showing the imbalance in the test set. In \cite{3}, they have achieved an accuracy of 83.75±3.6\% for ABIDE dataset and 76\% for DS030 dataset with precision, recall, F1-score as 74\%, 76\%, 72\% respectively for classification of good and average quality as "acceptable quality" and poor quality as "exclude". They extracted 64 IQMs for each sample and made use of methods such as SVM and RFC. Though it is not as computationally demanding as a deep learning model, the IQM calculations are still upto a certain extend computationally expensive. Using ABIDE and CombiRX dataset \cite{2} achieved an accuracy of 84\% with a recall of 77\% for classes '0' and '1'. Among all the methods, the highest accuracy of 95\% was achieved by \cite{21} using the CNN model with approximately 2 billion parameters for the a subset of 420 samples 1000BRAINS dataset and 19 in-house samples.

\begin{table}[h]
\centering
\begin{adjustbox}{max width=\textwidth}
\begin{tabular}{|l|c|c|c|p{5cm}|}
\hline
\textbf{Method} & \textbf{Dataset} & \textbf{Accuracy} & \textbf{\# Parameters} & \textbf{Remarks} \\
\hline
MRIQC (SVM+IQM)~\cite{3} & ABIDE & 83.75\% & -- & IQMs require extensive preprocessing \\
\hline
\makecell[l]{MIA~\cite{1} \\ (3D-CNN)} & MR-ART & 84.17\% & 220K & Deep CNN with multiple sites \\
\hline
\makecell[l]{IJNS~\cite{21} \\ (2D-CNN)} & 1000BRAINS & 95.00\% & 2B & High risk of overfitting \\
\hline
\textbf{Proposed (DHoGM)} & MR-ART & \textbf{94.34\%} & \textbf{209} & Lightweight and interpretable \\
\hline
\textbf{Proposed (DHoGM)} & ABIDE & \textbf{90.00\%} & \textbf{209} & Generalizes well across sites \\
\hline
\end{tabular}
\end{adjustbox}
\caption{Comparison with existing methods on MR-ART and ABIDE datasets.}
\label{tab:comparison}
\end{table}

\begin{table}[h!]
\centering
\renewcommand{\arraystretch}{1.0}
\resizebox{\textwidth}{!}{ 
\begin{tabular}{|c|c|c|c|c|}
\hline
\textbf{Category} & \textbf{Sample 1} & \textbf{Sample 2} & \textbf{Sample 3} & \textbf{Sample 4} \\
\hline
\raisebox{-.5\height}{\textbf{Samples}} & \includegraphics[width=2.5cm]{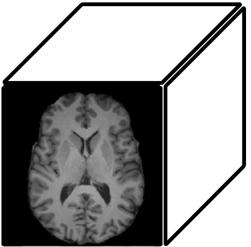} & \includegraphics[width=2.5cm]{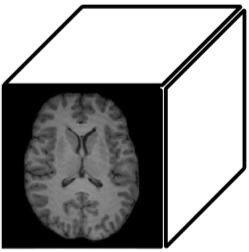} & \includegraphics[width=2.5cm]{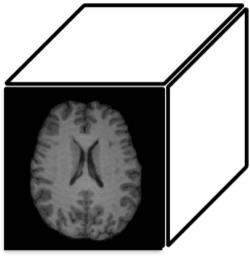} & \includegraphics[width=2.5cm]{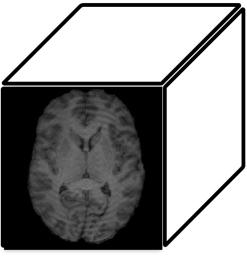} \\
\hline
\textbf{2D-DHoGM Result} & 1   & 1 & 2 & 2 \\
\hline
\textbf{2D-DHoGM Confidence} & 
\makecell{\(P_{2D,C1}\) = 0.8516 \\ \(P_{2D,C2}\) = 0.1484} & 
\makecell{\(P_{2D,C1}\) = 0.5012\\ \(P_{2D,C2}\) = 0.4988} & 
\makecell{\(P_{2D,C1}\) = 0.6896\\ \(P_{2D,C2}\) = 0.3104} & 
\makecell{\(P_{2D,C1}\) = 0.3765\\ \(P_{2D,C2}\) = 0.6235} \\
\hline
\textbf{3D-DHoGM Result} &  1 & 2 & 1 & 2 \\
\hline
\textbf{3D-DHoGM Confidence} & 
\makecell{\(P_{3D,C1}\) = 0.5387 \\ \(P_{3D,C2}\) = 0.4613} & 
\makecell{\(P_{3D,C1}\) = 0.6489\\ \(P_{3D,C2}\) = 0.3511} & 
\makecell{\(P_{3D,C1}\) = 0.8855\\ \(P_{3D,C2}\) = 0.1145} & 
\makecell{\(P_{3D,C1}\) = 0.1876\\ \(P_{3D,C2}\) = 0.8124} \\
\hline
\textbf{Final Result} & 1 & 2 & 2 & 2 \\
\hline
\textbf{Final Confidence} & 
\(P_1\) = 0.6951& 
\(P_2\) =  0.4249& 
\(P_2\) = 0.2124& 
\(P_2\) = 0.7179\\
\hline
\end{tabular}
}
\caption{Final class predictions and confidence}
\label{tab:comparison_of_OPs}
\end{table}
 
\begin{figure}[h!]
\centering
\includegraphics[width=0.45\textwidth]{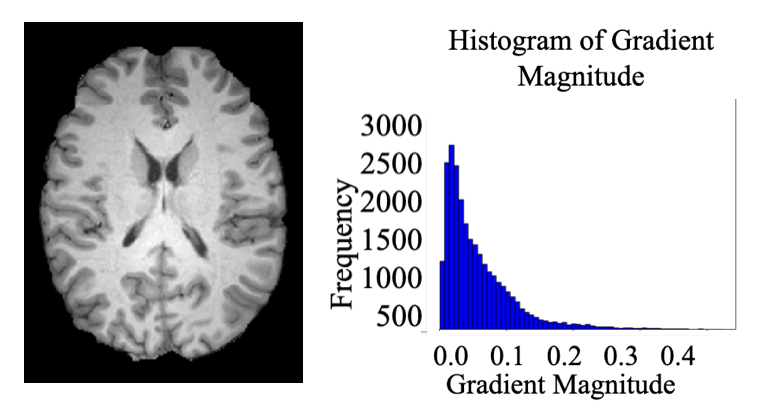}
\caption{Decision of model based on DHoGM metric changes at Gaussian noise with mean = 0 and standard deviation of 0.0103 with PSNR = 39.41 dB}
\label{fig:Noise_Decision}
\end{figure}

\item{Qualitative Analysis : } Table \ref{tab:comparison_of_OPs} displays the prediction and the confidence score of each of the models and the final prediction along with the final confidence score. The decision made by our model based on the proposed metric DHoGM changes at Gaussian noise with mean = 0 and standard deviation = 0.0103 with PSNR = 39.41 dB as displayed in Figure \ref{fig:Noise_Decision}. The noise is not visually recognizable but the metric is sensitive to noise below 40 dB approximately, showing the efficiency of our proposed method.

\item{Ablation: Impact of Feature Strategy}
We conducted an ablation study to evaluate the individual and combined contributions of the two parallel strategies within our framework. When using only the 2D DHoGM-based MLP model, the system achieved an accuracy of 90.12\%, while the improved 3D DHoGM-based approach, leveraging threshold-based classification, reached accuracy of 91.45\% exclusively. When both strategies were fused, the system achieved a significantly higher accuracy of 94.34\%, validating the effectiveness of our combined approach. Given the complementary strengths of the two models and the rationale already discussed, these results empirically justify our decision to implement a strict logical AND-type fusion: a sample is classified as "poor quality" if either model identifies it as such. This decision rule reinforces the robustness of our pipeline by prioritizing sensitivity to quality issues over partial agreement, and it aligns with our goal of conservative and reliable quality control.


\item{Runtime and Efficiency}
Our model uses only 209 trainable parameters and requires less than 50 seconds per subject on a TPU backend with approximately 5.3 GB of RAM and 20.2 GB of disk space consumed during runtime, with an average processing time of 54 seconds per sample, making it well-suited for real-time deployment and scalable integration into clinical and research workflows.
\end{enumerate}


\section{Discussion}

We proposed an interpretable method for detecting motion artifacts in T1-weighted brain MRI using gradient-based features. Our dual-path framework integrates slice-level (2D) and volume-level (3D) DHoGM features to capture both localized and global motion-induced degradation. The approach demonstrated strong performance on both in-domain (MR-ART) and out-of-domain (ABIDE) datasets, achieving up to 94.3\% accuracy with only 209 trainable parameters. 
Compared to deep learning models requiring millions of parameters, our method provides a favorable trade-off between performance, efficiency, and interpretability. It avoids extensive preprocessing and training requirements, enabling faster deployment and easier integration into clinical workflows. The HoGM-based feature extraction captures structural distortions caused by motion, and the dual-path strategy enhances detection sensitivity without sacrificing specificity. 
Our ablation experiments showed that combining 2D and 3D features significantly improves classification accuracy over either pathway alone, validating the complementary nature of local and global analysis. The model’s conservative fusion rule, flagging a scan as poor quality if either stream detects motion ensures high sensitivity, which is critical for clinical quality control. 
From a translational perspective, the method supports rapid, site-independent screening of MRI quality and can be integrated into automated pipelines to flag scans for rescanning or manual review. The low parameter count and minimal runtime make it scalable for large datasets. 
This study has several limitations. While cross-site performance was demonstrated on ABIDE, further validation is needed across diverse scanner types, populations (e.g., pediatrics), and acquisition protocols. The thresholding mechanism, while effective, could be replaced with calibrated probabilistic models. Lastly, our current binary classification could be extended to multi-grade motion scoring.
In summary, the proposed method offers an effective, interpretable, and scalable solution for automated motion artifact detection in brain MRI, with strong potential for deployment in clinical and research workflows.

\newpage


\begin{thebibliography}{99}

\bibitem{1} 
P. Vakli, B. Weiss, J. Szalma, et al.; Automatic brain {MRI} motion artifact detection based on end-to-end deep learning is similarly effective as traditional machine learning trained on image quality metrics; Medical Image Analysis , Volume 88, August 2023, 102850

\bibitem{2} 
S J Sujit, I. Coronado, A. Kamali, P A Narayana , R E Gabr; Automated image quality evaluation of structural brain MRI using an ensemble of deep learning networks; Journal of Magnetic Resonance Imaging, Volume50, Issue4, October 2019, Pages 1260-1267

\bibitem{3} 
O. Esteban, D. Birman, M. Schaer, O. O. Koyejo, R. A. Poldrack, and K. J. Gorgolewski; Mriqc: Advancing the automatic prediction of image quality in mri from unseen sites; PloS one

\bibitem{4} 
 I. Fantini, C. Yasuda, M. Bento, L. Rittner, F. Cendes, and R. Lotufo; Automatic mr image quality evaluation using a deep cnn: A reference-free method to rate motion artifacts in neuroimaging; Computerized Medical Imaging and Graphics,  Jun:90:101897

\bibitem{5}
K. Meding; A. Loktyushin; M. Hirsch; Automatic detection of motion artifacts in MR images using CNNS; International Conference on Acoustics, Speech, and Signal Processing (ICASSP), 2379-190X

\bibitem{6}
D Romascano, M Rebsamen, P Radojewski, et al.; Cortical thickness and grey-matter volume anomaly detection in individual MRI scans: Comparison of two methods; NeuroImage: Clinical, Volume 43, 2024, 103624

\bibitem{7} 
Á. Nárai , P. Hermann1, T. Auer , P. Kemenczky1, J. Szalma1, I. Homolya1, E. Somogyi1, P. Vakli1, B. Weiss1, Z. Vidnyánszky; Movement-related artefacts (MR-ART) dataset of matched motion-corrupted and clean structural MRI brain scans; Scientific Data

\bibitem{8} 
Di Martino, Adriana and Yan, Chao-Gan and Li, Qingyang and Denio, Erin and Castellanos, Francisco X and Alaerts, Kaat and Anderson, Jeffrey S and Assaf, Michal and Bookheimer, Susan Y and Dapretto, Mirella and others; The autism brain imaging data exchange: towards a large-scale evaluation of the intrinsic brain architecture in autism; Molecular psychiatry

\bibitem{9}
P. Kaur and A. K. Sao; Gradient profile based super resolution of MR images with induced sparsity; Medical Image Computing and Computer Assisted Intervention–MICCAI 2018, (pp.109-117)

\bibitem{10}
P. Kaur, A. K. Sao, and C. K. Ahuja; Super resolution of magnetic resonance images;  Journal of Imaging, 2021, 7(6), 101


\bibitem{11}
J. M. Slipsager, S. L. Glimberg, L. Højgaard, R. R. Paulsen, P. Wighton, M. D. Tisdall, C. Jaimes, B. A. Gagoski, P. E. Grant, A. van Der Kouweet al.; Comparison of prospective and retrospective motion correction in 3d-encoded neuroanatomical mri; Magnetic resonance in medicine, 2022 Feb;87(2):629-645

\bibitem{12}
Andersson, J. L. R., \& Sotiropoulos, S. N.; Motion artifacts in MRI: A complex problem with many partial solutions; Journal of Magnetic Resonance Imaging, 2015

\bibitem{13}
T. Budrys, V. Veikutis, S. Lukosevicius, R. Gleizniene , E. Monastyreckiene, I. Kulakiene; Artifacts in magnetic resonance imaging: how it can really affect diagnostic image quality and confuse clinical diagnosis?; Journal of Vibroengineering, . MAR 2018, VOL. 20, ISSUE 2. ISSN 1392-8716 

\bibitem{14}
Y. Chang, Z. Li, G. Saju, H. Mao, and T. Liu; Deep learning-based rigid motion correction for magnetic resonance imaging: a survey; MetaRadiology, Volume 1, Issue 1, June 2023, 100001


\bibitem{15}
I. Havsteen, A. Ohlhues , K. H. Madsen, J. D. Nybing , H. Christensen, A. Christensen; Are Movement Artifacts in Magnetic Resonance Imaging a Real Problem?—A Narrative Review; Frontiers in Neurology, Volume 8 - 2017

\bibitem{16}
B J Casey et al., The Adolescent Brain Cognitive Development (ABCD) study: Imaging acquisition across 21 sites; Developmental Cognitive Neuroscience
Volume 32, August 2018, Pages 43-54

\bibitem{17}

D.C. Van Essen et al., The Human Connectome Project: A data acquisition perspective, Volume 62, Issue 4, 1 October 2012, Pages 2222-2231

\bibitem{18}
K. S. Nayak, D. G. Nishimura, A. Macovski; Magnetic Resonance Imaging: Principles and Techniques; Journal of Clinical Imaging Science, 2015


\bibitem{19}
D. G. Norris; Principles of Magnetic Resonance Assessment of Brain Function; Journal of Magnetic Resonance Imaging, 2006


\bibitem{20} 
N Nithianandam, P Kaur, A K Sao; A Simple yet Effective Method for Motion Detection in Structural Magnetic Resonance Images; IEEE International Symposium on Biomedical Imaging

\bibitem{21} 
E. Roecher, L. MCosch, J. Zweerings, el al.; Motion artifact detection for t1-weighted brain mr images using convolutional neural networks; International Journal of Neural Systems

\bibitem{22}
B. Kocak, et al.,Evaluation metrics in medical imaging AI: fundamentals, pitfalls, misapplications, and recommendations; European Journal of Radiology Artificial Intelligence
Volume 3, September 2025, 100030
  
\bibitem{23}
Y. Zhan and R. Zhang; No-reference image sharpness assessment based on maximum gradient and variability of gradients; IEEE Transactions on Multimedia, Volume: 33 Issue: 1


\bibitem{24}
K. Khurshid, K. L. Berger, R. J. McGough; Automated PET/CT brain registration for accurate attenuation correction; 2009 Annual International Conference of the IEEE Engineering in Medicine and Biology Society  

\bibitem{25}
J. P. Hornak; The Basics of MRI; Interactive Learning Software, Rochester Institute of Technology


\bibitem{26}
Firbank MJ, Coulthard A, Harrison RM, Williams ED; Partial volume effects in MRI studies of multiple sclerosis; Magnetic Resonance Imaging, 1999


  
\bibitem{27}
Thomas Küstner, Annika Liebgott, Lukas Mauch, Petros Martirosian, Fabian Bamberg, Konstantin Nikolaou, Bin Yang, Fritz Schick , Sergios Gatidis; Automated reference-free detection of motion artifacts in magnetic resonance images; Computerized Medical Imaging and Graphics, 
Volume 90, June 2021
  
\bibitem{28}
A. Krizhevsky, I. Sutskever, G. E. Hinton; ImageNet Classification with Deep Convolutional Neural Networks; Advances in Neural Information Processing Systems , Volume 60, Issue 6
Pages 84 - 90

\bibitem{29}
P. Kaur, J S Thornton, F Barkhof, T A. Yousry, S Vos, H Zhang; Quality assessment of {MR} images: Does deep learning outperform machine learning with handcrafted features on new sites?; Proceedings of the International Society for Magnetic Resonance in Medicine (ISMRM).

\bibitem{30}
Bertelsen, A., et al.; MRI-only treatment planning: benefits and challenges; Physics in Medicine and Biology, 2018


\bibitem{31}
UNC Health Appalachian; Comparing MRI, CT, and PET Scans: How They Work and When They're Used; UNC Health Appalachian Blog, 2024


\bibitem{32}
Kellner, E., Dhital, B., \& Reisert, M.; Gibbs-Ringing Artifact Removal Based on Local Subvoxel-shifts; arXiv preprint arXiv:1501.07758, 2015



\bibitem{33} 
V.P.B Grover, J M Tognarelli, M M E Crossey, el al.; Magnetic Resonance Imaging: Principles and Techniques: Lessons for Clinicians; J Clin Exp Hepatol, 2015, 5(3):246–255. 

\bibitem{34}
N. Dalal and B. Triggs; Histograms of oriented gradients for human detection; 2005 IEEE Computer Society Conference on Computer Vision and Pattern Recognition 

\bibitem{35}
E. Bourigault, A. Hamdi, A. Jamaludin; X-Diffusion: Generating Detailed 3D MRI Volumes From a Single Image Using Cross-Sectional Diffusion Models; arXiv preprint

\bibitem{36}
D.Tamada; Noise and artifact reduction for MRI using deep; arXiv preprint arXiv:2002.12889, 2020


\bibitem{37}
 F. Kruggel; A simple measure for acuity in medical images; IEEE Transactions on Image Processing, Volume: 27 Issue: 11,

\bibitem{38}
P. Kaur, A. Singh Minhas, C. K. Ahuja, and A. Kumar Sao; Estimation of 3T MR images from 1.5T images regularized with physics based constraint; Medical Image Computing and Computer Assisted Intervention – MICCAI 2023, pp 132–141

\bibitem{39}
S. Kastryulin, J. Zakirov, N. Pezzotti, and D. V. Dylov; Image quality assessment for magnetic resonance imaging ; IEEE Access,  V.11 pp. 14154-14168, 2023

\bibitem{40}
J Eugenio Iglesias, G Lerma-Usabiaga, L Carlos Garcia-Peraza-Herrera, S Martinez ; Retrospective Head Motion Estimation in Structural Brain MRI with 3D CNNs ; MICCAI 2017

\bibitem{41}
A. Loktyushin, C. Schuler, B. Schölkopf; Retrospective Motion Correction of Magnitude-Input MR Images; Medical Learning Meets Medical Imaging, 2015
pp 3–12

\bibitem{42}
C. Jaimes, D. J. Murcia, K. Miguel, C. DeFuria, P. Sagar, and M. S. Gee; Identification of quality improvement areas in pediatric mri from analysis of patient safety reports; Pediatric Radiology, Volume 48, pages 66–73, (2018)

\bibitem{43}
 P. Kaur, A. K. Sao, and C. K. Ahuja; Unsupervised reconstruction of high-field-like magnetic resonance images from low-field magnetic resonance images regularized with magnetic resonance image-based priors; Signal, Image and Video Processing, volume 16, issue 5, pages 1313-1321

\bibitem{44}
R. Fluss, D. Faraggi, and B. Reiser; Estimation of the Youden Index and its Associated Cutoff Point; Biometrical Journal 47 (2005) 4, 458–472

\bibitem{45}
Jeffrey P. Woodard \& Monica P, Carley-Spencer, No-Reference image quality metrics for structural MRI, 

\end{thebibliography}
\end{document}